\documentclass[11pt]{article}
\usepackage{moriond,epsfig}

\def\be{\begin{equation}}
\def\ee{\end{equation}}
\def\bea{\begin{eqnarray}}
\def\eea{\end{eqnarray}}

\def\apjs{ApJS} % Astrophysical Journal, Supplement
\def\mnras{MNRAS} % Monthly Notices of the RAS
\def\aap{A\&A} % Astronomy and Astrophysics
\def\be#1{\begin{equation}\label{#1}}
\def\ee{\end{equation}}

\begin{document}
\vspace*{4cm}
\title{BLIND MD-MC COMPONENT SEPARATION FOR POLARIZED OBSERVATIONS OF THE CMB WITH THE EM ALGORITHM}

\author{ J. AUMONT }

\address{Laboratoire de Physique Subatomique et de Cosmologie \\
 53, avenue des Martyrs, Grenoble, France}

\maketitle\abstracts{We present the {\sc PolEMICA} \cite{aumont} (Polarized
Expectation-Maximization Independent Component Analysis) algorithm
which is an extension to polarization of the SMICA \cite{delabrouille} temperature component separation method.
This algorithm allows us to estimate blindly in harmonic space multiple physical 
components from multi-detectors
polarized sky maps. Assuming a linear noisy mixture of components we
are able to reconstruct jointly
the electromagnetic spectra of the components for each mode $T$, $E$ and $B$, as well as
the temperature and polarization spatial power spectra, $TT$, $EE$, $BB$, $TE$, $TB$ and $EB$ for 
each of the physical components and for the noise on each of the
detectors.
This has been tested using full sky simulations of the Planck satellite polarized channels for
a 14-months nominal mission assuming a simple linear sky model including CMB, and optionally
Galactic synchrotron and dust emissions.}

%+++++++++++++++++++++++++++++++++++++++++++++++++++++++++++++++++++++++++++++
%+++++++++++++++++++++++++++++++++++++++++++++++++++++++++++++++++++++++++++++
%+++++++++++++++++++++++++++++++++++++++++++++++++++++++++++++++++++++++++++++
%+++++++++++++++++++++++++++++++++++++++++++++++++++++++++++++++++++++++++++++
%+++++++++++++++++++++++++++++++++++++++++++++++++++++++++++++++++++++++++++++

\section[]{Introduction}
\label{intro}

Mapping the Cosmic Microwave Background (CMB) polarization is one
of the major challenges of future missions in observational cosmology.
CMB polarization is linear and therefore can be described by the first three Stokes
parameters I, Q and U which are generally combined to produce three fields
(modes), $T$, $E$ and $B$.
The polarization of the CMB photons carries extra physical informations that are not
accessible by the study of the temperature anisotropies. Therefore its
measurement helps breaking down the degeneracies on cosmological
parameters as encounter with temperature anisotropies measurements
only. Furthermore, the study of the CMB
polarization is also a fundamental tool to estimate the energy scale
of inflation.  

However, CMB polarization is several orders of magnitude weaker than the
temperature signal and therefore, its detection needs an efficient
separation between the CMB and the astrophysical foregrounds which are
expected to be significantly polarized.

A direct subtraction of these foreground contributions on the CMB data
will require an accurate knowledge of their spatial distributions and
of their electromagnetic spectra. But these latter are not yet well characterized
in polarization.    

To try to overcome the above limitations, a great amount of work has
been dedicated to design and implement algorithms
for component separation which can discriminate between CMB and
foregrounds. We present here the {\sc PolEMICA} (Polarized
Expectation-Maximization Independent Component Analysis) algorithm
which is an extension of the Spectral Matching
Independent Component Analysis (SMICA) \cite{delabrouille} which 
has been developed to consider both a fully blind analysis for which
no prior is assumed and a semi-blind analysis incorporating previous
physical knowledge on the astrophysical components. This extension
allows to estimate jointly the temperature and polarization parameters
from a set of multi-frequencies $I$, $Q$ and $U$ sky maps.

%+++++++++++++++++++++++++++++++++++++++++++++++++++++++++++++++++++++++++++++
%+++++++++++++++++++++++++++++++++++++++++++++++++++++++++++++++++++++++++++++
%+++++++++++++++++++++++++++++++++++++++++++++++++++++++++++++++++++++++++++++
%+++++++++++++++++++++++++++++++++++++++++++++++++++++++++++++++++++++++++++++
%+++++++++++++++++++++++++++++++++++++++++++++++++++++++++++++++++++++++++++++

\section[]{Model of the microwave and sub-mm sky}\label{simus}

\indent To perform the separation between CMB and the astrophysical foregrounds, the diversity of the electromagnetic
spectra and of the spatial spectra of the different components is generally used. 
Observations from a multi-band instrument, for the Stokes parameters $I$, $Q$ and
$U$, can
be modeled as a linear combination of multiple physical components
leading to what is called a Multi-Detectors Multi-Components (MD-MC) modeling.

\indent  Assuming an experiment with $n_\nu$
detector-bands at frequencies $\nu_i$ and $n_c$ physical components
in the data, working in the spherical harmonics space, we can model
the observed sky for $X=\{T,E,B\}$, for each frequency band and for
each $\{\ell,m\}$

\be{data}
y^{\nu,X}_{\ell m}=\sum_{c=1}^{n_c}A^{\nu,X}_cs^{c,X}_{\ell
  m}+n^{\nu,X}_{\ell m}
\ee

\noindent where $y^{\nu,X}_{\ell m}$ is a vector of size $(3\cdot
n_\nu\cdot n_\ell\cdot n_m)$ containing the observed data, $s^{c,X}_{\ell m}$ is a $(3\cdot
n_c\cdot n_\ell\cdot n_m)$ vector describing each component template and $n^{\nu,X}_{\ell m}$ is a vector
of the same size than $y^{\nu,X}_{\ell m}$ accounting for the noise. $A^{\nu,X}_c$
is the \emph{mixing matrix} containing the electromagnetic behaviour
of each component and is of size $(3\cdot n_\nu)\times(3\cdot n_c)$

The aim of the component separation algorithm presented in here
is to extract $A^{\nu,X}_c$, $s^{c,X}_{\ell m}$ and $n^{\nu,X}_{\ell m}$ 
from the $y^{\nu,X}_{\ell m}$ sky observations.

%+++++++++++++++++++++++++++++++++++++++++++++++++++++++++++++++++++++++++++++
%+++++++++++++++++++++++++++++++++++++++++++++++++++++++++++++++++++++++++++++
%+++++++++++++++++++++++++++++++++++++++++++++++++++++++++++++++++++++++++++++
%+++++++++++++++++++++++++++++++++++++++++++++++++++++++++++++++++++++++++++++
%+++++++++++++++++++++++++++++++++++++++++++++++++++++++++++++++++++++++++++++

\section[]{A MD-MC component separation method for polarization}\label{mdmc}

To reduce the number of unknown parameters in the model 
described by equation (\ref{data}), it is interesting to 
rewrite this equation in terms of the temperature and polarization auto and
cross power spectra and to bin them over $\ell$ ranges.

\be{dataspec}
R_y(b) = AR_s(b)A^T+R_n(b)
\ee

\noindent where $R_y(b)$ and $R_n(b)$ are
$(n_\nu\cdot3)\times(n_\nu\cdot3)$ matrices and $R_s(b)$ is a
$(n_c\cdot3)\times(n_c\cdot3)$ matrix. We assume that the physical components in the data are statistically
independent and uncorrelated and that the noise is uncorrelated between channels.

To estimate the above parameters from the data we have extended
to the case of polarized data the spectral matching algorithm
developed in SMICA \cite{delabrouille} for temperature only.
The key issue of this method is to estimate these parameters, or some of them  (for a semi-blind analysis),
by finding the best match between the model density matrix, $R_y(b)$,
computed for the set of estimated parameters
and the data density matrix $\tilde{R}_y(b)$ obtained from
the multi-channel data. The likelihood function is a reasonnable
measure of this mismatch. We have extended this method to jointly deal
with the temperature and polarization power spectra and also to
estimate the $TE$, $TB$ and $EB$ cross power spectra \cite{aumont}.

The maximization of the likelihood function is achieved via the
Expectation-Maximization algorithm (EM) \cite{em}. This algorithm
will process iteratively from an initial value of the parameters following a sequence of parameter updates called `EM steps'. By construction each EM step improves the spectral fit by maximizing
the likelihood. For a more detailed review of the spectral matching EM
algorithm used here, see \cite{delabrouille}. 

%+++++++++++++++++++++++++++++++++++++++++++++++++++++++++++++++++++++++++++++
%+++++++++++++++++++++++++++++++++++++++++++++++++++++++++++++++++++++++++++++
%+++++++++++++++++++++++++++++++++++++++++++++++++++++++++++++++++++++++++++++
%+++++++++++++++++++++++++++++++++++++++++++++++++++++++++++++++++++++++++++++
%+++++++++++++++++++++++++++++++++++++++++++++++++++++++++++++++++++++++++++++

\section[]{Simulated microwave and sub-mm sky as seen by Planck}

Following the MD-MC model discussed above and given an observational setup, 
we construct, using the HEALPix pixelization
scheme \cite{healpix} and in CMB temperature
units, fake $I$, $Q$ and $U$ maps of the sky at each of the instrumental frequency bands.
For these maps we consider three main physical components
in the sky emission: CMB, thermal dust and synchrotron. Instrumental noise
is modeled as white noise.
 
The CMB component map is randomly generated from the
polarized CMB angular power spectra
for a set of given cosmological parameters. In the following we have used $H_0=71\ {\rm
  km}\cdot{\rm s}^{-1}\cdot{\rm Mpc}^{-1}$, $\Omega_b=0.044$,
$\Omega_m=0.27$, $\Omega_\Lambda=0.73$ and $\tau=0.17$ that are the
values of the cosmological concordance model according to the WMAP 1 year results
 \cite{wmappc}. 

For the diffuse Galactic synchrotron emission we use the 
template maps in temperature and in polarization provided by
\cite{giardino}. Here we have chosen to use a constant spectral index equal to the mean of the spectral index map, $\alpha=-2.77$,
so that the simple linear model of the data holds.        

In the case of the thermal dust we dispose of few 
observational data of the polarized diffuse emission and to date no template 
for this is available. 
Thus, we have considered a power-law model, renormalized to mimic at large angular scales
the $TE$ cross power spectrum measured by Archeops at 353~GHz \cite{tearcheops}.
$I,Q$ and $U$ full-sky maps are generated randomly from
these power spectra. We extrapolate them to each of the frequency of interest by assuming
a grey body with an emissivity of 2.

Noise maps for each channel are generated from white noise realizations
normalized to the nominal level of instrumental noise for that channel. \\

We have performed sets of simulations 
of the expected Planck satellite data to intensively test the algorithm presented above.
We present here results from 300 realizations considering full-sky
maps at the LFI and HFI polarized channels,
30, 40 and 70~GHz for LFI and 100, 143, 217 and 353~GHz for HFI for a nominal
14-month survey. We have simulated maps at
$n_{\rm side}$ = 512 which permits the
reconstruction of the angular power spectra up to $\ell\simeq1500$. 
The reconstructed spectra will be
averaged over bins of size 20 in $\ell$.

%+++++++++++++++++++++++++++++++++++++++++++++++++++++++++++++++++++++++++++++
%+++++++++++++++++++++++++++++++++++++++++++++++++++++++++++++++++++++++++++++
%+++++++++++++++++++++++++++++++++++++++++++++++++++++++++++++++++++++++++++++
%+++++++++++++++++++++++++++++++++++++++++++++++++++++++++++++++++++++++++++++
%+++++++++++++++++++++++++++++++++++++++++++++++++++++++++++++++++++++++++++++

\section[]{Results}

We have applied the {\sc PolEMICA} component separation algorithm
to the simulations presented above. From them, we have computed the data density matrix $R_{y}$ and applied the
algorithm. We simultaneously estimate the $R_s$, $R_n$
and $A$ matrices, with no priors, for temperature and polarization. To ensure the reliability of the
results we have performed 10000 EM iterations and checked, for each
simulation, the convergence of the EM algorithm.

\begin{figure}
\centering
 \includegraphics[width=17cm,height=10cm]{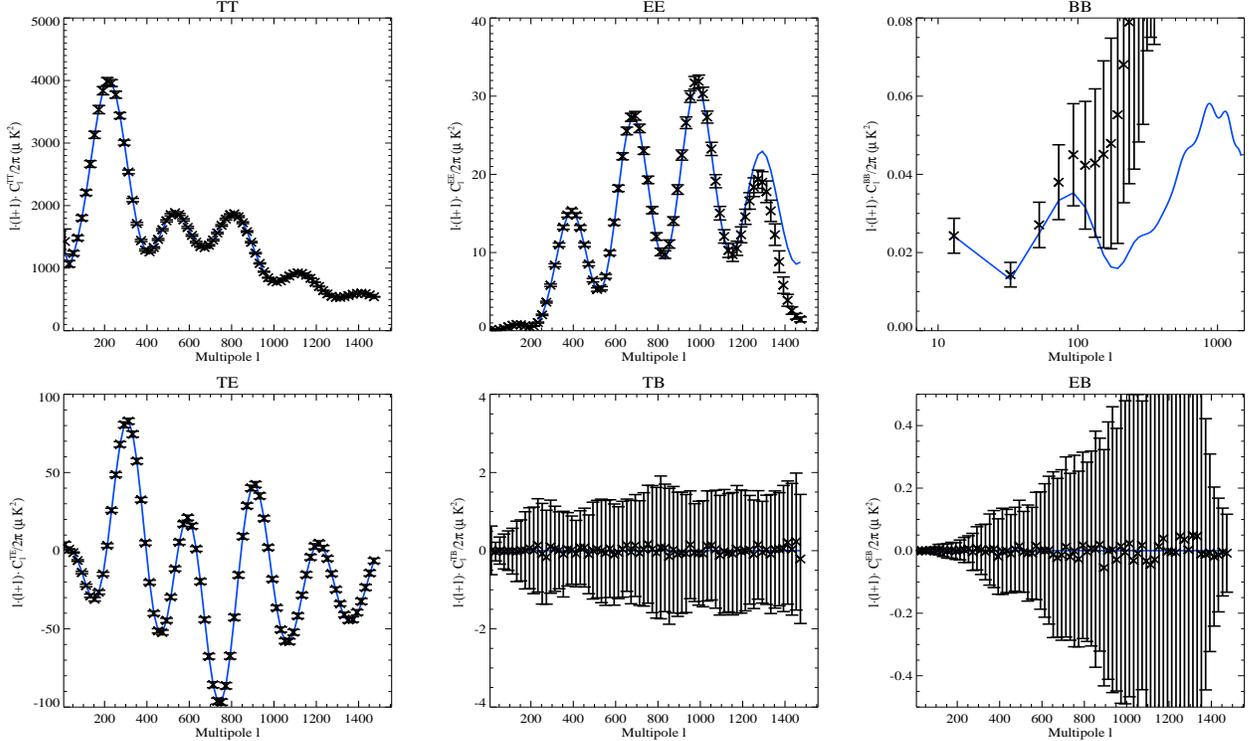}
\caption{Reconstruction of the CMB power spectra for
$C_\ell^{TT}$, $C_\ell^{EE}$, $C_\ell^{BB}$, $C_\ell^{TE}$,
$C_\ell^{TB}$ and $C_\ell^{EB}$ at 100 GHz in $\mu {\rm K}_{\rm CMB}^2$
performed on Planck full sky maps simulations. Crosses
represent the reconstructed spectra, solid lines the input model. Error bars are the dispersion over
$\sim$ 300 simulations.}\label{fig:smallscales}
\end{figure}

We present in figure \ref{fig:smallscales} the reconstructed CMB power
spectra. We can see that for the $TT$, $TE$, $TB$ and $EB$ spectra, we
are able to reconstruct the $C_\ell$ over the full range of $\ell$
values that are accessible at this pixelization resolution ($\ell_{\rm
  max}\sim1500$). The $EE$ spectrum is recovered accurately up
to $\ell\simeq1200$. For smaller angular scales, a bias appears. This
bias is a pixelization problem that would occur at a larger $\ell$ if
the resolution was higher. The $BB$ spectrum is reconstructed up to
$\ell\simeq70$. For larger $\ell$, the reconstructed spectrum is
residual noise arising from the fact that the convergence of the
EM algorithm is slow and therefore we have not properly
converged. This bias appears in our separation when the signal over
noise ratio is below $10^{-2}$ and does not affect the reconstruction
of the other parameters. Even if we were able to avoid this effect,
the recovered $BB$ spectrum would be compatible with zero for
$\ell>70$ thanks to the size of the error bars.

The power spectra from our input synchrotron and dust emissions are
recovered with efficiency up to $\ell\simeq1500$ for $TT$, $EE$, $BB$,
$TE$, $TB$ and $EB$ \cite{aumont}. Power spectra of the noise in temperature and in
polarization are also fully reconstructed \cite{aumont}. 

The mixing matrix $A$ elements corresponding to CMB and dust emission
are recovered efficiently, for temperature and polarization. For the
synchrotron emission, mixing matrix elements corresponding to
polarization are well recovered and those corresponding to temperature
are biased at intermediate frequency values \cite{aumont}. This bias
is due tu a slight mixing up between synchrotron and CMB in
temperature. It does not happen in polarization where the synchrotron
dominates the CMB. This bias can be avoided by the adjunction of
priors in the separation, like for example assuming an equal electromagnetic
spectrum in temperature and polarization for each component \cite{aumont}

To evaluate the impact of foregrounds in the determination of the CMB temperature and polarization
power spectra we have compared the results of the presented analysis
to those on simulations that contain only CMB and noise.
In the presence of foregrounds, the error bars on the reconstruction of the CMB power spectra 
are increased by at least a factor of two both in temperature and in polarization \cite{aumont}.
Therefore, although the foreground contribution in the data can be removed,
it significantly reduces the precision to which the CMB polarization
signal can be extracted from the data.


\begin{thebibliography}{}

\bibitem{aumont}
Aumont~J. \& Mac\'{\i}as-P\'erez~J.-F., 2007, MNRAS, in press, {\tt astro-ph/0603044}

\bibitem{delabrouille}
Delabrouille~J., Cardoso~J.-F. \& Patanchon~G., 2003, \mnras, 346, 1089

\bibitem{em}
Dempster~A., Laird~N. \& Rubin~D, 1977, J. of the Roy. Stat. Soc. B, 39, 1

\bibitem{giardino}
Giardino~G., Banday~A.~J., G\'orski~K.~M., Bennet~K., Jonas~J.~L. \& Tauber~J., 2002, \aap, 387, 82

\bibitem{healpix}
G\'orski~K.~M., Hivon~E. \& Wandelt~B.~D., 1999, {\tt astro-ph/9812350}

\bibitem{tearcheops}
Ponthieu~N. et al., 2005, \aap, 444, 327

\bibitem{wmappc}
Spergel~D.~N. et al., 2003, \apjs, 148, 175

\end{thebibliography}
\end{document}